\documentclass[%
 reprint,
 amsmath,amssymb,
 aps,
]{revtex4-2}

\usepackage{graphicx}% Include figure files
\usepackage{dcolumn}% Align table columns on decimal point
\usepackage{bm}% bold math
\usepackage{color}
\begin{document}

\preprint{APS/123-QED}

\title{Unraveling complex magnetism in two-dimensional FeS$_2$}% Force line breaks with \\

\author{Duo Wang}
\email{Duo.Wang@physics.uu.se}
\affiliation{Dept. of Physics and Astronomy, Uppsala University, 75120 Uppsala, Sweden}
\author{Xin Chen}
\email{cchenxinn@gmail.com}
\affiliation{Dept. of Physics and Astronomy, Uppsala University, 75120 Uppsala, Sweden}
\author{Biplab Sanyal}
\email{Biplab.Sanyal@physics.uu.se}
\affiliation{Dept. of Physics and Astronomy, Uppsala University, 75120 Uppsala, Sweden}

% \email{Second.Author@institution.edu}
%\affiliation{%
% Authors' institution and/or address\\
% This line break forced with \textbackslash\textbackslash
%}%

%\collaboration{MUSO Collaboration}%\noaffiliation

\date{\today}% It is always \today, today,
             %  but any date may be explicitly specified

\begin{abstract}
	Atomically thin two dimensional magnets have given rise to emergent phenomena due to magnetic exchange and spin-orbit coupling showing a great promise for realizing ultrathin device structures. In this paper, we critically examine the magnetic properties of 2D FeS$_2$, which has been recently claimed to exhibit room temperature ferromagnetism in the (111) orientation. Our ab initio study based on collinear density functional theory has revealed the ground state as an antiferromagnetic one with an ordering temperature of around 100K along with a signature of spin-phonon coupling, which may trigger a ferromagnetic coupling via strain. Moreover, our calculations based on spin-spirals indicate the possibility of non-collinear magnetic structures, which is also supported by Monte Carlo simulations based on ab initio magnetic exchange parameters. This opens up an excellent possibility to manipulate magnetic structures by the application of directional strain.
\end{abstract}

%\keywords{Suggested keywords}%Use showkeys class option if keyword
                              %display desired
\maketitle

%\tableofcontents

\section{INTRODUCTION}
Magnetism in two dimension (2D) has long been at the heart of numerous theoretical, experimental and technological advances as this research area has been proven to be a fertile ground for emergent magnetic phenomena, such as the study of topology\cite{doi:10.1021/acs.jpclett.7b00222, PhysRevB.102.115162, 10.1126/sciadv.aaw5685}, multiferroicity\cite{PhysRevLett.125.017601,C9CP06966F,10.1038/s41467-019-10693-0,C7NR09588K}, proximity effects in heterostructures\cite{10.1126/science.aav4450} \textit{et al}. Theoretically, long-range magnetic order is prohibited in the 2D isotropic-Heisenberg model at a finite temperature by the Mermin-Wagner theorem. However, symmetry breaking, such as the occurrence of magnetic anisotropy, removes this restriction. The discovery of CrI$_3$, a 2D material that retains ferromagnetic ordering in a monolayer\cite{10.1038/nature22391} has opened a new era of tunable magnetism in the ultra-thin scale. A number of 2D magnets have emerged after 2017, such as Cr$_2$Ge$_2$Te$_6$\cite{10.1038/nature22060}, Fe$_3$GeTe$_2$\cite{10.1038/s41563-018-0149-7}, CrX$_3$ (X=Cl, Br)\cite{C9CP01837A, 10.1002/admi.201901410}, MnSe$_2$\cite{10.1021/acs.nanolett.8b00683}, MPS$_3$ (M=Ni, Mn, Co)\cite{PhysRevB.98.134414, C8TC05011B, 10.1088/1361-648x/aa8a43}, CrTe$_2$\cite{10.1007/s12274-020-3021-4}, and Cr$_2$Te$_3$\cite{10.1021/acs.nanolett.9b05128}. The weak van der Waals bonding between layers in the bulk 3D magnetic systems offers a unique advantage to realize 2D or quasi 2D forms of these systems. However, the challenge is to have a magnetic long range order and thermal stability at room temperature and above.

\par Bulk FeS$_2$ is a paramagnetic material with a band gap around 0.9 eV, which makes it a promising candidate for applications in the field of photovoltaic and photoelectron chemical cells\cite{Ennaoui_1985, 10.1016/0379-6787(84)90009-7, 10.1063/1.4996551}. FeS$_2$ is similar to MnSe$_2$\cite{10.1021/acs.nanolett.8b00683}, which has been demonstrated to have high-temperature ferromagnetism down to monolayers whereas their bulk counterparts are low-temperature antiferromagnetic materials. In the past, dynamical stability, electronic and magnetic properties of 2D FeS$_2$ with different surface orientations have been investigated theoretically\cite{10.1016/s0039-6028(02)01849-6, 10.1016/s0039-6028(02)02294-x, 10.1021/jp100578n, 10.1016/j.susc.2013.08.014, 10.1016/j.commatsci.2015.01.035}. Very recently, Anand \textit{et al} have reported the isolation of 2D FeS$_2$, and more importantly, they found that it is possible to observe long-range ferromagnetic (FM) order even up to room temperature\cite{2010.03113, 10.1021/acs.jpcc.1c04977}. Their DFT simulations show that the FeS$_2$ (111) surface is the one that has the lowest formation energy and at the same time has FM configuration as the ground state. However, the occurence of room temperature ferromagnetism is quite intriguing and calls for critical analysis. 

\par In this paper, our aim is to investigate the properties of FeS$_2$ (111) monolayer in detail. Firstly, we found that the structure with antiferromagnetic (AFM) configuration is more stable than the FM structure, the energy difference being quite small and sensitive to the crystal structure, which indicates that there is a substantial spin-lattice coupling. Furthermore, we studied the interatomic magnetic exchange parameters and ordering temperatures for AFM and FM structural phases. The results show that the ordering temperatures are 102 K and 32 K for crystal structures with AFM and FM coupling respectively. Moreover, spin-spiral calculations utilizing generalized Bloch's theorem indicate that the magnetic ground state of the (111) surface is actually non-collinear, which is also consistent with our results obtained from Monte Carlo simulations where canting of magnetic moments deviated from collinear magnetic configuration is observed.

\section{COMPUTATIONAL METHODS}
Our computational approach is based on density functional theory\cite{DFT} using plane wave basis set and projector augmented wave method\cite{PAW, PAW2} implemented in the Vienna Ab initio Simulation Package (VASP)\cite{VASP}. The plane wave energy cutoff was set to 600 eV. A $11 \times 11 \times 1$ k-mesh was used for integration in the Brillouin zone, and the forces on atoms were converged to 0.001 eV/\AA.
The optimized structures were used as input for the full-potential linearized muffin-tin orbital (FP-LMTO) code RSPt\cite{rspt} to calculate the magnetic anisotropy energy (MAE) and interatomic magnetic exchange parameters. The calculations of interatomic exchange parameters were based on magnetic force theorem (MFT)\cite{MFT}, which maps the magnetic excitations onto the Heisenberg Hamiltonian:
\begin{equation}
    \hat{H}=-\sum_{i \neq j}J_{ij} \Vec{e_{i}} \cdot \Vec{e_{j}},
\end{equation}
where $J_{ij}$ is the interatomic exchange interaction between the two magnetic moments at sites \textit{i} and \textit{j}, e is the unit vector along the magnetization direction at site \textit{i} (or \textit{j}). %The idea behind this procedure is that an infinitesimal rotation of two magnetic moments at two sites in a collinear magnetic ground state will cause an energy variation in the Heisenberg model proportional to the exchange parameter and two angles.
It has following form
\begin{equation}
	J_{ij}^{m_1m_2}=\frac{T}{4}\sum_{-\infty}^{\infty} [\widetilde{\hat{\Delta}}_{i,m_1} \cdot G_{ij,m_1m_2}^{\uparrow}\cdot \widetilde{\hat{\Delta}}_{j,m_2}\cdot G_{ji,m_2m_1}^{\downarrow}],
\end{equation}
in which $\hat{G}_{ij}$ is real-space intersite Green's function, and $\hat{\Delta}_i$ is the exchange splitting on site $i$. Theoretically, it is possible to obtain orbitally decomposed exchange parameter for the system in any basis since both $\hat{G}_{ij}$ and $\hat{\Delta}_i$ are orbitally dependent. However, the above method has not been implemented yet in the code and therefore, we can only roughly get `t$_{2g}$-like' and `e$_g$-like' states by diagonalizing the intersite $J_{ij}$ matrix.
The k-mesh in the Brillouin zone was performed using $31 \times 31 \times 1$ points.
To describe the exchange-correlation effects, we used the generalized gradient approximation (GGA)\cite{gga} augmented by the Hubbard-U correction (GGA+U$_{eff}$)\cite{PhysRevB.57.1505}. The Coulomb $U_{eff}$ (U-J) value we used for Fe-d electrons is 2 eV, which has been proven to be a good value for 2D Fe structures\cite{2010.03113,10.1038/s41565-018-0134-y}. In our simulations, we took Fe \textit{3s3p3d4s} and S \textit{3s3p} electrons as valence electrons.
Finally, we used the extracted $J_{ij}$ and MAE to calculate the magnetic ordering temperature by means of classical Monte Carlo (MC) simulations for the solution of the Heisenberg Hamiltonian, and also adiabatic magnon spectra as implemented in the Uppsala Atomistic Spin Dynamics (UppASD)\cite{uppasd} code. For this purpose, a 70$\times$70$\times$1 (19600 atoms) cell was considered. The transition state barriers are obtained by using the climbing image Nudged Elastic Band (cNEB) method\cite{henkelman2000climbing}.

\section{RESULTS}
\begin{figure}[htbp]
\includegraphics[scale=1.0]{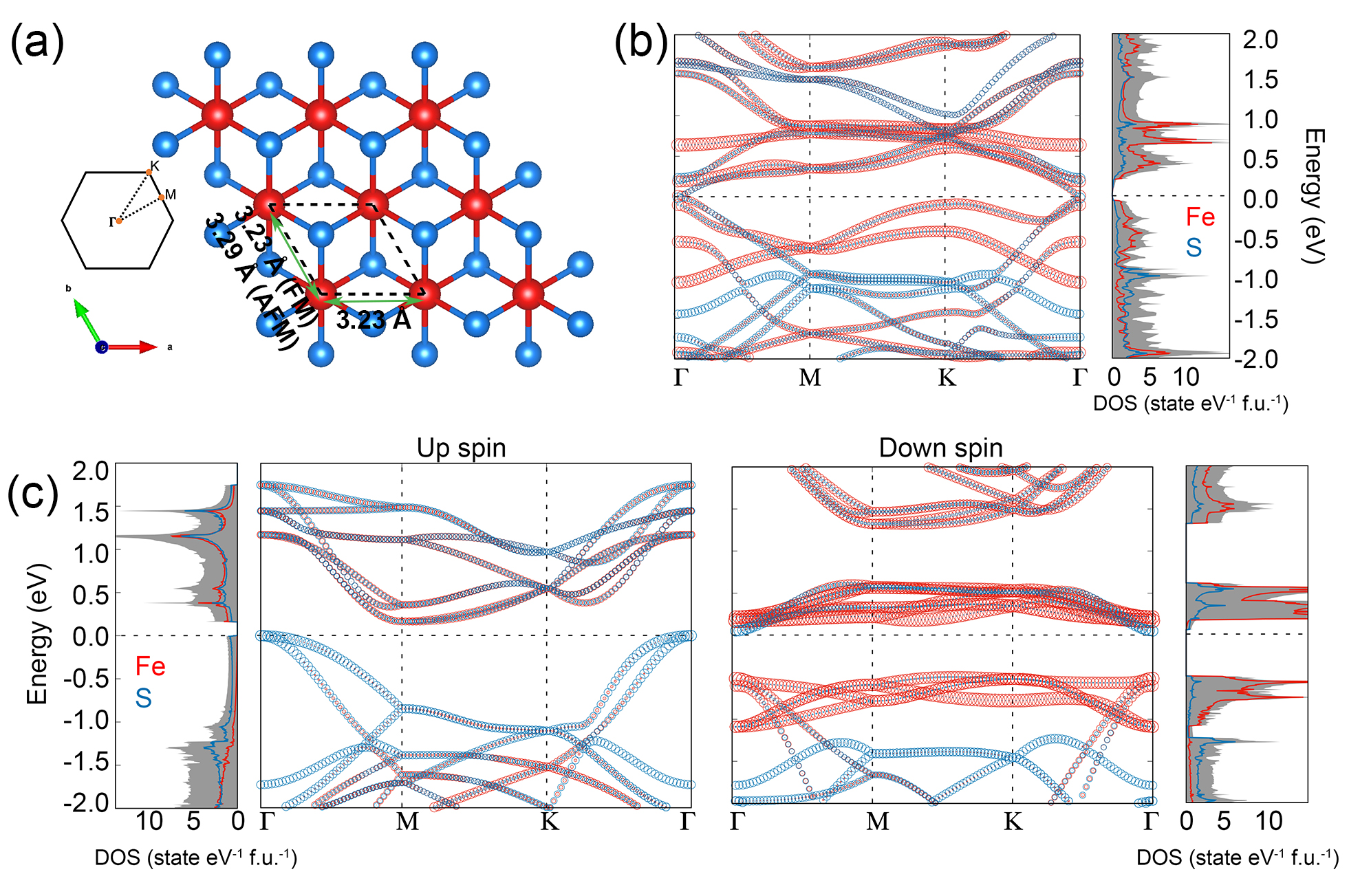}
\caption{(a) The top view of (111) surface. Fe and S atoms are indicated as blue and red balls respectively. The unit cell is shown in the dashed line, and Fe-Fe distances are marked with green arrows. The first Brillioun zone with high symmetry points is shown. Density of states (DOS) and atom-projected band structure for up-spin and down-spin electrons of AFM structure (b) and FM structure (c) are shown. Total DOS (shaded regions) and partial DOS of Fe (red curves) and S (blue curves) are shown. In the projected band structures, the size of circles is proportional to the weight of the atomic contribution. The Fermi level is set to 0 eV. }
\label{fig1}
\end{figure}

The FeS$_2$ (111) surface was considered due to its lowest formation energy and dynamical stability\cite{2010.03113, 10.1016/j.commatsci.2015.01.035}. The crystal structure belongs to the family of layered transition metal dichalcogenides, in which two S layers sandwich the Fe layer. There are six nearest neighbor Fe atoms around each Fe atom, forming a hexagonal close-packed structure. Each Fe atom (nominal charge state as Fe$^{4+}$) with six nearest neighbor S atoms form an octahedral structure offering a crystal field splitting of d-states into t$_{2g}$ (3 electrons) and e$_g$ (1 electron) states. 
Our accurate calculations of total energies (confirmed by both plane wave PAW and FP-LMTO methods) indicate that the FeS$_2$ (111) surface with AFM configuration has lower energy compared to the FM state, which is in contrast to some previous studies\cite{2010.03113,10.1016/j.commatsci.2015.01.035}. However, the energy difference between these two phases is relatively small (7.2 meV/atom) and is sensitive to the crystal structure. The top view of these two structures are shown in Fig.\ref{fig1}(a). The FM structure is symmetrical with all the next nearest neighbor Fe-Fe distances are exactly the same whereas for the structure with AFM Fe-Fe coupling, asymmetry appears, and the Fe-Fe distance along the AFM coupling direction is longer than the direction with FM coupling. This magnetic coupling dependent Fe-Fe distance indicates the presence of spin-lattice coupling. Practically, it means that the magnetic configuration is tunable by introducing external strain along a specific direction. It also implies that the ferromagnetism observed in recent experiments might have been triggered by strain. However, the observation of room-temperature ferromagnetism is still elusive as our calculated Curie temperature for a symmetric ferromagnetic structure is very low, which has been discussed later. It should be mentioned that in addition to the (111) surface, we also studied the (100) and (110) surfaces. Our results show that when the thickness is small, the structure with AFM configuration always shows lower energy than the FM case. Detailed description and data are shown in the supplementary material \cite{SuppMater}.
%============================================
%============================================
\par The calculated band structures are shown in Figs.\ref{fig1}(b) and \ref{fig1}(c). The FM structure has a very small band gap $\sim$ 50 meV at the $\Gamma$ point between spin up valence band maximum with dominating S character and spin down conduction band minimum with dominating Fe character. Moreover, both spin channels have their respective band gaps (0.16 and 0.55 eV for the spin up and spin down channel). It is interesting to note that electron (hole) doping will give rise to predominantly spin-down (spin-up) conduction. The AFM structure shows a semimetallic behavior with occupied and unoccupied bands meeting at the $\Gamma$ point. It should be noted that a relatively high DOS at the Fermi level for the FM structure compared to the vanishing DOS for the AFM structure reflects the relative stability of the AFM structure. Our calculated MAEs of these structural phases show that the easy axis is along the [100] direction for the AFM structure, and energy differences are around 3.40 meV/Fe compared to the structure with other magnetization directions ([001], [010], and [110]). For the FM case, due to higher symmetry, the easy axes are along the [100] and [010] directions, and the MAE is 3.27 meV/Fe, which is at the same level as the AFM case.

%============================================
\begin{figure}[hbt!]
\includegraphics[scale=1]{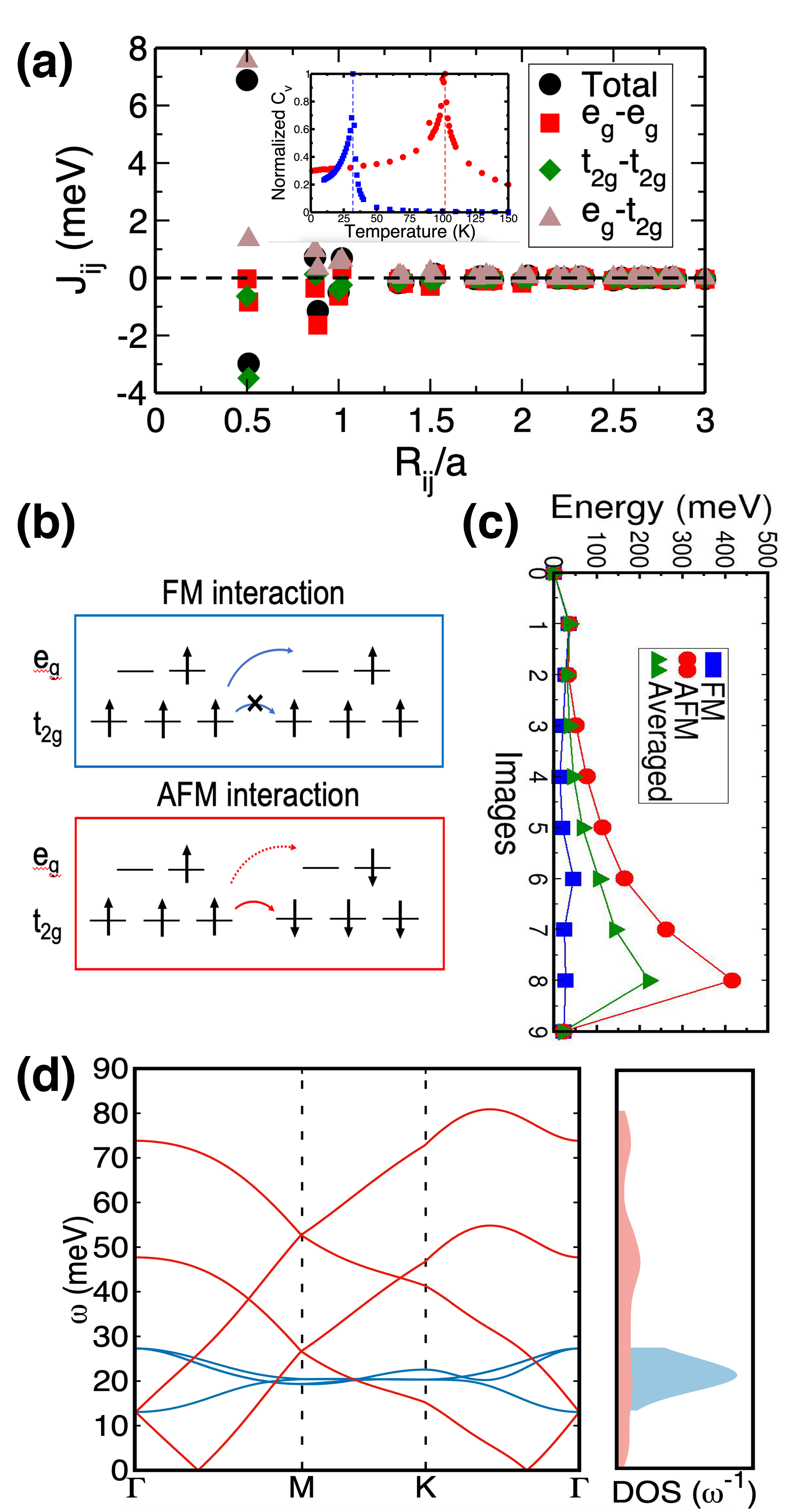}
\caption{(a) Calculated total and orbital-resolved exchange parameters of Fe-Fe atoms as a function of distance. The positive and negative values indicate ferromagnetic and antiferromagnetic couplings, respectively. Inset shows the normalized specific heat as a function of temperature, calculated from Classical Monte Carlo simulations. (b) Schematic picture of the orbital dependent interactions. (c) Calculated energy barriers from AFM structure to FM structure. For both crystal structures, results for FM and AFM alignments are shown along with the average values.  (d) Adiabatic magnon spectra together with magnon DOS for FM (blue) and AFM (red) structures.}
\label{fig2}
\end{figure}
%============================================
\par Calculated total and orbitally-resolved exchange parameters of the AFM structure are shown in Fig.\ref{fig2}(a). Each Fe-Fe pair has two exchange parameters due to different couplings along different directions. As seen from Fig. \ref{fig2}(a), the strength of couplings decreases fast as the Fe-Fe distance increases. Only the first three nearest neighbors play an essential role, and the most significant contribution comes from the nearest neighbor couplings. There are six Fe atoms around each Fe atoms as the first magnetic nearest neighbor; two of them show FM coupling and the other four show AFM coupling. The magnitude of the FM coupling is two times bigger than the AFM one (6.88 and -2.98 meV, respectively). The physical mechanism is understood from orbitally decomposed exchange parameters. As mentioned above, there are four electrons in the d orbital of each Fe atom, three of them are located at the t$_{2g}$ level, and the other one is at the e$_g$ level. Fig.\ref{fig2}(b) shows a schematic picture of the different exchange interactions. Hopping of the form e$_g$-like - t$_{2g}$-like (We will call them t$_{2g}$ and e$_g$ in the latter to keep the context simple) leads to an exchange coupling that is predominantly FM because of the local Hund's coupling. On the other hand, the t$_{2g}$-t$_{2g}$ hopping is prohibited for FM alignment (shown in blue box), whereas this is allowed for AFM alignment (shown in red box). Therefore, t$_{2g}$-t$_{2g}$ hybridization leads to AFM coupling. Another hopping of the form e$_g$-e$_g$ changes between AFM and FM as Fe-Fe distance increases, by contrast, the strength of this coupling is close to zero and much smaller than the other two and is therefore negligible. For example, the nearest neighbor ferromagnetic exchange parameter is 6.88 meV, of which e$_g$-t$_{2g}$ and t$_{2g}$-t$_{2g}$ contributions are 7.55 and -0.63 meV respectively. For the AFM coupling, the contributions come from e$_g$-t$_{2g}$ and t$_{2g}$-t$_{2g}$ as 1.34 and -3.49 meV respectively. All of these Fe-Fe exchange interactions are mediated by the hybridization by the S-p orbitals. The structure-dependent magnetism originates from the dominated form in the coupling. For comparison, we also calculated the exchange parameters and magnetic ordering temperature for the FM structure. Results are shown in Fig. S1. Unlike the AFM structure, the total J$_{ij}^{'}$s of the FM structure are much smaller due to strong competition between different coupling forms. For the first nearest neighbor coupling, contributions from e$_g$-t$_{2g}$ and t$_{2g}$-t$_{2g}$ are 3.73 and -2.68 meV respectively. As a result, the total exchange parameter is only 0.36 meV.
%============================================
\begin{figure}[hbt!]
\includegraphics[scale=0.8]{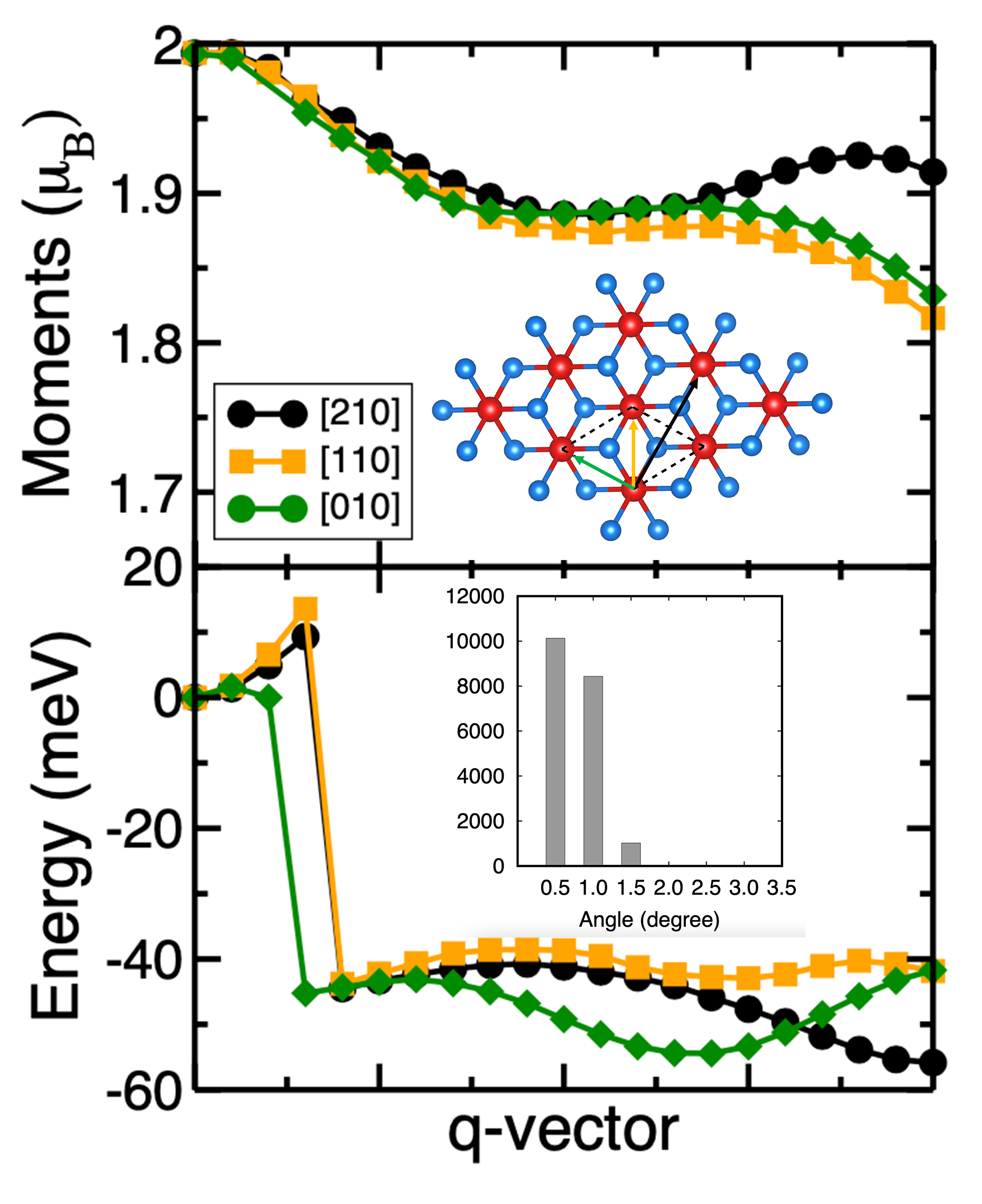}
\caption{Calculated magnetic moments (upper panel) and relative spin spiral energies (lower panel) as a function of q-vector for the AFM structure where the energy at q=[0,0,0] is set to zero. Inset in the upper panel shows crystal structure with spin-spiral propagation directions [010] (green), [110](red), [210] (black), which correspond to the curves in the upper and lower panel, and the unit cell is shown by the dashed lines. Inset in the lower panel shows the histogram of the deviation angles of the magnetic moments relative to the easy axis, obtained from MCS.}
\label{fig3}
\end{figure}
%============================================
\par In order to estimate the energy involved in the switching between FM and AFM alignments, we have calculate the energy barriers (shown in Fig. \ref{fig2}(c)) by nudged elastic band method. It is observed that the AFM alignments involves a much large energy barrier compared to the FM alignment. The average values of these two alignments are shown to have a rough estimate, which turns out to be $\sim$ 200 meV. This indicates that it is possible to switch from the ground state AFM to the FM structure by strain, which might have caused ferromagnetism in the recent experiment.
\par Furthermore, we performed classical Monte Carlo simulations (MCS) based on the calculated MAE values and exchange parameters. As shown in the inset in Fig. \ref{fig2}(a), the magnetic ordering temperature of AFM structure is 102 K whereas, for the FM structure, it is 32 K, which reflects the difference in total magnetic exchange parameters between the two.
\par Finally, based on the calculated exchange parameters and MAE, we calculated adiabatic magnon spectra for both AFM and FM structures as shown in Fig. \ref{fig2}(d). Firstly, there are 4 magnon branches for both structures, which is dictated by the number of Fe atoms in the unit cell. Secondly, the energy gaps at the $\Gamma$ point are around 13 meV for both AFM and FM structure due to the similar MAEs. Thirdly, the dispersion relation around the $\Gamma$ point is quadratic (linear) dependence for the FM (AFM) structure. At last, as it is shown in the magnon DOS, compared to the AFM structure, the magnon spectrum of the FM structure is more localized.
 
\par Based on the Generalized Bloch theorem, the spin spiral total energies can be calculated using the primitive cell. As shown in the inset of upper panel in Fig. \ref{fig3}, three arrows represent propagation directions of the spin-spiral vector but have been translated into real space. Magnetic moments and relative energies as a function of propagation vector are shown in the upper and lower panels of Fig. \ref{fig3}. The left and right endpoints represent FM and AFM state respectively. For the spin-spiral propagations along [010] and [110] direction, the AFM state has lower energy than the FM state, which is the same as the result we got from the collinear spin calculations. Remarkably, the magnetic ground state along these two directions is not AFM but spin-spiral states (q$_{[010]}$=[0.47, -0.23, 0], q$_{[110]}$=[0.23, 0.23, 0]). For another propagation direction ([210], black curve), the ground state occurs at the right endpoint-the AFM state (q$_{[210]}$=(0, 0.5, 0)).
\par The results for the FM structure are shown in Fig. S2. One difference is obvious if we compare it to the AFM structure: the spin spiral total energy as a function of q-vector in two directions ([010] blue, and [110] red) are exactly the same due to the higher crystal symmetry. The distances of all the first nearest Fe-Fe pairs are identical, so the spin spiral states in these two directions are also identical. These results, on the other hand, demonstrate the strong spin-lattice coupling in FeS$_2$. The magnetic ground state along these two directions is FM, which corresponds to our collinear spin calculations. For the states along another propagation direction ([210], the AFM configuration has the lowest energy. The stability of non-collinear magnetic arrangement obtained from our T=0K DFT calculations is also confirmed by the magnetic configurations obtained by Monte Carlo simulations at T=0.02K for the FM and AFM structures. The histogram of the deviation angle of the moments from the easy axis is shown in the inset in Fig.\ref{fig3} and Fig. S2. It is clearly seen that the magnetic moments are deviated from the direction of easy magnetization ([100] direction). %Secondly, the canting in the FM structure is more pronounced than the canting in the AFM structure, which indicates the instability of the FM structure.

%============================================
%\begin{figure}[htbp]
%\includegraphics[scale=0.13]{figure4.jpg}
%\caption{Simulated adiabatic magnon spectrum for FeS$_2$ (111) surface, AFM structure.}
%\label{figure4}
%\end{figure}
%============================================
\section{CONCLUSION}
\par In summary, we did systematic investigations on the magnetic properties of FeS$_2$ (111) monolayer. We found that the structure with AFM configuration has lower energy than the structure with FM configuration with the presence of spin-lattice coupling. The recently reported room-temperature ferromagnetism in experiments may have occurred due to sufficient strain in the samples. We calculated the magnetic exchange parameters and critical temperature for both AFM and FM phases, due to the different mechanisms of the exchange coupling and the competition between orbital channels, the magnetic ordering temperatures for these two phases are 102 and 32 K, respectively. Our calculated spin spiral energies using generalized Bloch theorem and also supported by the results from Monte Carlo simulations indicate that the magnetic ground states of the ground state structural phase are not perfectly collinear. The complex magnetism in FeS$_2$ (111) monolayer, tunable by strain may open up interesting applications in spin-driven devices.
\par B.S. acknowledges financial support from the project grant (2016-05366) and the Swedish Research Links programme grant (2017-05447) from Swedish Research Council. Also, Carl Tryggers Stiftelse is acknowledged for providing grant (CTS 20:378) to B.S. Duo Wang thanks the China scholarship council for financial support (No. 201706210084). B.S. and D.W. gratefully acknowledge supercomputing time allocation by the Swedish National Infrastructure for Computing (project no. SNIC2020-3-26) and PRACE DECI-15 project DYNAMAT.

\footnotesize During the revision of this paper, the experimental paper are published in JPCC\cite{10.1021/acs.jpcc.1c04977}. The authors revised the results as they point out that they observed the ferromagnetic phase, but it only appears at low temperature, which is consistent with our theoretical studies.

%\bibliography{reference}% Produces the bibliography via BibTeX.
%apsrev4-2.bst 2019-01-14 (MD) hand-edited version of apsrev4-1.bst
%Control: key (0)
%Control: author (8) initials jnrlst
%Control: editor formatted (1) identically to author
%Control: production of article title (0) allowed
%Control: page (0) single
%Control: year (1) truncated
%Control: production of eprint (0) enabled
\providecommand{\noopsort}[1]{}\providecommand{\singleletter}[1]{#1}%\providecommand{\noopsort}[1]{}\providecommand{\singleletter}[1]{#1}%
\end{document}